\documentclass[10pt,letterpaper]{article}
\usepackage{opex3}
\usepackage{graphicx}

\begin{document}

\title{Heisenberg limited Sagnac interferometry}

\author{Aziz Kolkiran and G S Agarwal}
\address{Department of Physics, Oklahoma State University,
Stillwater, OK - 74078, USA}
%\date{\today}
\email{aziz.kolkiran@okstate.edu}
\begin{abstract}
    We show how the entangled photons produced in parametric down conversion
    can be used to improve the sensitivity of a Sagnac
    interferometer. Two-photon and four-photon coincidences increases the
    sensitivity by a factor of two and four respectively. Our results apply to sources
    with arbitrary pumping and squeezing parameters.
\end{abstract}

\section{Introduction}\label{intro}
When two electromagnetic waves counter-propagate along a circular
path in rotation they experience different travel times to
complete the path. This induces a phase shift between the two
counter-propagating waves proportional to the angular velocity of
the rotation. This phase difference is called as the Sagnac effect
\cite{Sagnac 1913a} and in addition to its scientific importance,
it has numerous practical applications such as detection and
high-precision measurement of rotation. It was studied and used in
optics only with lasers until the new work \cite{Bertocchi 2006}
where they demonstrated the single-photon interference in the
fiber Sagnac interferometer using spontaneous parametric down
conversion as the source of single photons. However, it turns out
that the results of interference are no different than with
classical sources. This is also true of many interferometric
experiments done at the single photon level \cite{Grangier 1986,
Hariharan 1993, Zeilinger 2000}. Thus a natural question would be
--what is the nature of interference if we replace the single
photon source by entangled photon pair source. This is what we
examine in detail. We find that the sensitivity of Sagnac
interferometer could be considerably improved by using correlated
photons \cite{Holland 1993, Dowling 1998}. We thus bring Sagnac
interferometer in the same class as other experiments on imaging
\cite{Abouraddy 2001,Pittman 1995}, lithography \cite{Boto
2000,Yablonovitch 1999,Korobkin 2002,Agarwal 2001,Bjork
2001,Dangelo 2001} and spectroscopy \cite{Agarwal 2003}.
%In particular, the use of photon pairs in interferometers allows
%phases to be measured to the precision in the Heisenberg limit
%where uncertainty scales as $1/N$ \cite{Giovannetti 2004} as
%compared to the shot noise limit $1/\sqrt{N}$.

 Parametric down conversion (PDC) is predominant mechanism for
experimentalists to create entangled photon pairs as well as
single photons. Multi-photon entangled states produced in the
down-conversion process is often used in quantum information
experiments and applications like quantum cryptography and the
Bell inequalities. In particular, demonstrations of two-photon
\cite{Ou 1990a, Rarity 1990, Ou 1990b, Edamatsu 2002} and
four-photon \cite{Eibl 2003, Walther 2004} interferences are
holding promise for realizable applications with
entanglement-enhanced performance. The principle of this
enhancement lies in the fact that ``the photonic de Broglie
wavelength" \cite{Jacobson 1995} of an ensemble of photons with
wavelength $\lambda$ and number of photons $n$ can be measured to
be $\lambda/n$ using a special interferometer. Further Steuernagel
\cite{Steuernagel 2002} has proposed the measurement of the
reduced de Broglie wavelength of two- and four-photon wave
packets.

In this paper we present an analysis of how parametric down
converted photons could be useful to increase the rotation
sensitivity in Sagnac interferometers. The results show two- and
four-fold increase in the sensitivity which can be interpreted as
a sign of two- and four-photon interference effect.
%The experiment
%\cite{Bertocchi 2006} with single photons leads us to think that
%demonstrations with the entangled photon pairs should be feasible
%with current technology and that Sagnac interferometers can be
%operated at much higher level of precision.
The organization of the paper is as follows. The Sagnac ring
interferometer is described in section \ref{sagnacinterferometer}
and the Sagnac phase shift is derived. In section
\ref{sagnacwithclassandquanth}, we analyze interference results
with classical and quantum inputs. We compare the results obtained
from entangled photon pairs input with classical and single-photon
inputs. We show how the two-photon and four-photon coincidences
increases the sensitivity in the phase shift. The visibility of
the counts are also discussed. We conclude the paper in section
\ref{conclusion} with a brief discussion on the disturbances that
can effect the transmission of modes in fibers.

\section{The Sagnac interferometer} \label{sagnacinterferometer}
The Sagnac interferometer consists of a ring cavity around which
two laser light beams travel in opposite directions on a rotating
base. One can form an interference pattern by extracting and
heterodyning portions of the two counter-propagating beams to
detect the rotation rate of the ring cavity relative to an
inertial frame. The position of the interference fringes is
dependent on angular velocity of the setup. This dependence is
caused by the rotation effectively shortening the path distance of
one beams, while lengthening the other. In 1926, a Sagnac
interferometer has been used by Albert Michelson and Henry Gale to
determine the angular velocity of the Earth. It can be used in
navigation as a ring laser gyroscope, which is commonly found on
fighter planes, navigation systems on commercial airliners, ships
and spacecraft.
\begin{figure}[h]
  \centering
  \scalebox{1.6}{\includegraphics{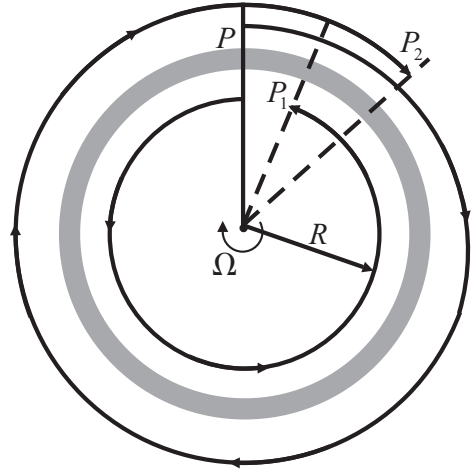}}
  \caption{Schematics of a Sagnac ring interferometer.  }
  \label{Fig1}
\end{figure}

The Sagnac effect \cite{Sagnac 1913a} can be understood by
considering a circular ring interferometer like the one shown in
Fig. \ref{Fig1}. The input laser field enters the interferometer
at point $P$ and split into clockwise (CW) and counterclockwise
(CCW) propagating beams by a beam splitter. If the interferometer
is not rotating, the beams recombine at point $P$ after a time
given by
\begin{equation}\label{rest time}
t=\frac{2\pi R}{c},
\end{equation}
where $R$ is the radius of the circular beam path. However, if the
interferometer is rotating with angular velocity $\Omega$, about
an axis through the center and perpendicular to the plane of the
interferometer, then the beams reencounter the beam splitter at
different times. The transit times to complete one round trip for
CW ($t_2$ at point $P_2$) and CCW ($t_1$ at point $P_1$) are given
by,
\begin{eqnarray}\label{transit times}
t_1=\frac{2\pi R}{c+R\Omega}\,\,,\\
t_2=\frac{2\pi R}{c-R\Omega}\,\,.
\end{eqnarray}
Then one round trip time delay between the two beams is the
difference
\begin{equation}\label{delay}
\Delta t=t_2-t_1=\frac{4\pi R^2 \Omega}{c^2-R^2\Omega^2}\,\,.
\end{equation}
For non-relativistic perimeter speeds (i.e. reasonable values of
$R$ and $\Omega$), $R^2\Omega^2 \ll c^2$, then
\begin{equation}\label{delay2}
\Delta t=\frac{4\pi R^2 \Omega}{c^2}\,\,.
\end{equation}
The angular phase difference between the two counter propagating
waves, the Sagnac effect, can be written as,
\begin{equation}\label{sagnac effect}
\phi=\omega\Delta t=\frac{8\pi}{\lambda\,c}A\Omega\,,
\end{equation}
where $\lambda$ is the wavelength, $c$ the light velocity in
vacuum, $A$ the interferometer area and $\Omega$ the angular
velocity of the interferometer. A more general approach \cite{Post
1967, Schleich 1984, Jacobs 1982} shows that the phase shift does
not depend on the shape of the interferometer and it is
proportional to the flux of the rotation vector $\Omega$ through
the interferometer enclosed area. Then one can increase the flux
by using multi-turn round-trip path like utilizing an optical
fiber. In terms of the total length of the optical fiber, $L$, we
can recast Eq. (\ref{sagnac effect}) into
\begin{equation}\label{sagnac effect 2}
\phi=\frac{4\pi LR\Omega}{\lambda\,c}\,.
\end{equation}
Eq. (\ref{sagnac effect 2}) shows that the phase shift induced by
rotation of a Sagnac fiber ring interferometer increases linearly
with the total length of the optical fiber.
\begin{figure}[h]
  \centering
  \scalebox{1.8}{\includegraphics{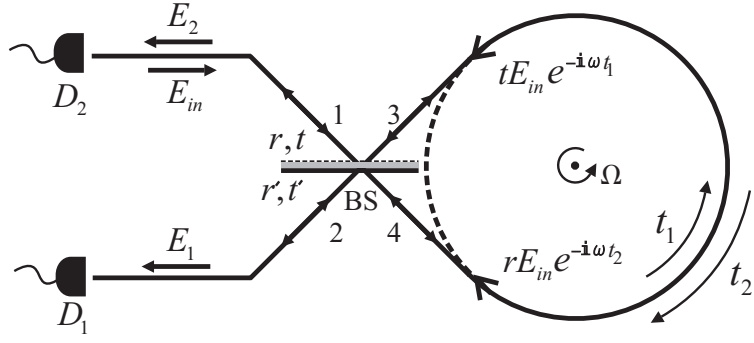}}
  \caption{The Sagnac interferometer setup with classical input. The input field $E_{in}$
  is separated by the beam splitter into two counter-propagating waves $tE_{in}$ and $rE_{in}$.
  Because of the rotation they end up at the beam splitter at different times ($t_1, t_2$).}
  \label{Fig2}
\end{figure}

\section{The Sagnac interferometer with classical and quantum
inputs}\label{sagnacwithclassandquanth}
%{\bf Classical Input}.
\subsection{Classical input}
We now consider a Sagnac fiber ring interferometer setup shown in
Fig. \ref{Fig2}. The two input ports 1 and 2 are mixed by a
$50/50$ beam splitter and sent through a rotating loop of fiber in
the opposite direction. Then the beams recombine at the beam
splitter and come out from the ports they entered. The rotation
induces the phase difference $\phi$ given by the Eq. (\ref{sagnac
effect 2}). If we choose the transmission and reflection
coefficients of the beam splitter as $t=1/\sqrt{2}=t'$,
$r=i/\sqrt{2}=r'$ then the entire setup transforms the input field
$E_{in}$ into the output fields $E_1$ and $E_2$ by
%\begin{equation}\label{transforms}
%\left(\begin{array}{c}E_1'\\E_2'\end{array}\right)=e^{-i(\omega
%t_2-\phi/2)}\left(\begin{array}{cc}i\sin(\phi/2) & \cos(\phi/2)
%\\ \cos(\phi/2) & i\sin(\phi/2)
%\end{array}\right)\left(\begin{array}{c}E_1\\E_2\end{array}\right),
%\end{equation}
\begin{eqnarray}\label{classical transform}
E_1&=&r'rE_{in}e^{-i\omega t_2}+t^2E_{in}e^{-i\omega t_1}=E_{in}e^{-i\omega t_2}ie^{i\phi/2}\sin(\phi/2),\\
E_2&=&rtE_{in}e^{-i\omega t_1}+t'rE_{in}e^{-i\omega
t_2}=E_{in}e^{-i\omega t_2}ie^{i\phi/2}\cos(\phi/2),
\end{eqnarray}
where $\omega$ is the frequency of the input field. The intensity
measurements at the detectors $D_1$ and $D_2$ becomes
\begin{eqnarray}
I_1&=&|E_1|^2=|E_{in}|^2\sin^2(\phi/2),\label{intensity1}\\
I_2&=&|E_2|^2=|E_{in}|^2\cos^2(\phi/2),\label{intensity2}
\end{eqnarray}
respectively.
\begin{figure}
  \centering
  \scalebox{1.8}{\includegraphics{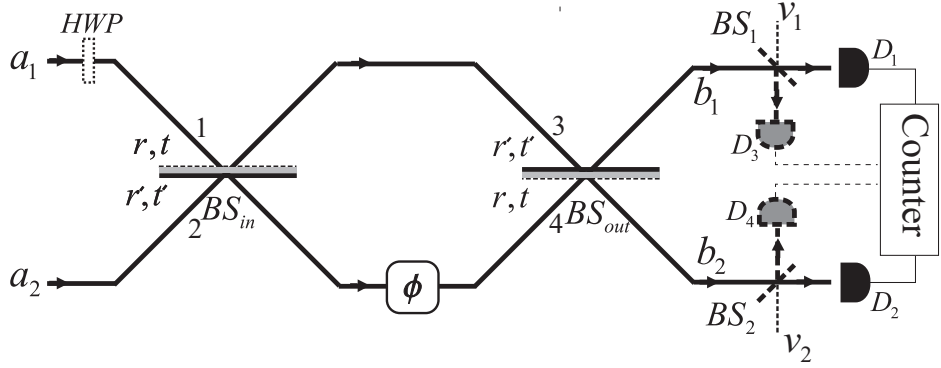}}
  \caption{The equivalent optical network diagram of the Sagnac interferometer for quantized
  fields. $``\,\phi\,"$ represents the phase shift provided by the rotating loop of the interferometer.
  The detectors $D_3$ and $D_4$ with the extra beam splitters (dashed lines) are to be used for four-photon coincidence
  counting.}
  \label{Fig3}
\end{figure}

\subsection{Quantum inputs}
%{\bf Quantum Inputs}.
Now we analyze the results with quantized fields. Figure
\ref{Fig3} shows the equivalent optical network diagram of the
interferometer. We denote $a_1$ and $a_2$ as the input mode
operators. The two beam splitters represent double-use of the
actual beam splitter. The output modes $b_1$ and $b_2$ are related
to the input modes by
\begin{eqnarray}\label{quantum transform}
\left(\begin{array}{c}b_1\\b_2\end{array}\right)&=&\underbrace{\frac{1}{\sqrt{2}}\left(\begin{array}{cc}1
& i\\ i & 1
\end{array}\right)}_{BS_{out}}\underbrace{\left(\begin{array}{cc}1& 0\\ 0 &
e^{i\phi}\end{array}\right)}_{SAGNAC}\underbrace{
\frac{1}{\sqrt{2}}\left(\begin{array}{cc}1 & i\\ i & 1
\end{array}\right)}_{BS_{in}}\left(\begin{array}{c}a_1\\a_2\end{array}\right),\nonumber\\
&=&ie^{i\phi/2}\left(\begin{array}{cc}-\sin(\phi/2) & \cos(\phi/2)\\
\cos(\phi/2) &
\sin(\phi/2)\end{array}\right)\left(\begin{array}{c}a_1\\a_2\end{array}\right),
\end{eqnarray}
where the global phase $ie^{i\phi/2}$ can be dropped. The use of
half-wave plate ($HWP$) is required when the input ports have
polarizations orthogonal to each other. Now the input and output
modes are related to each other by the linear transformation,
\begin{equation}\label{scattering matrix}
b_i=\sum^2_{j=1}S_{ij}a_j,
\end{equation}
where the matrix $S$ of the coefficients $S_{ij}$ is known as the
scattering matrix associated with the network. In fact, Eq.
(\ref{scattering matrix}) refers to the Heisenberg picture, where
the state vectors are constant while operators evolve. Therefore,
without knowing the Hamiltonian that describes the evolution by
the unitary operator $U$ on the state vectors, by using the
dynamics of the operators
\begin{eqnarray}
a_i\rightarrow b_i=\sum^2_{j=1}S_{ij}a_j\equiv U^{\dag}a_iU,\label{operator dynamics1}\\
a_i^{\dag}\rightarrow
b_i^{\dag}=\sum^2_{j=1}S^*_{ij}a_j^{\dag}\equiv
U^{\dag}a_i^{\dag}U,\label{operator dynamics2}
\end{eqnarray}
one can calculate the probabilities for detecting certain number
of photons at certain outputs.

Now, let us analyze the rotation sensitivity to the phase shift
$``\,\phi\,"$ for some Fock state inputs. We denote $n$-photons in
mode $a_1$ and $m$-photons in mode $a_2$ by $|nm\rangle$. First,
we begin with the input state $|10\rangle$, that is a single
incident photon in mode $a_1$ with the other mode in vacuum state.
The output state can be written as
\begin{equation}\label{single photon output}
U|10\rangle=Ua_1^{\dag}|00\rangle=Ua_1^{\dag}U^{\dag}U|00\rangle=Ua_1^{\dag}U^{\dag}|00\rangle.
\end{equation}
The last equality results from the fact that the interferometer
has no effect on the vacuum $|00\rangle$. Although we are in the
Schr\"odinger picture, it is perfectly valid to use Eq.
(\ref{operator dynamics2}) with the substitution $U\rightarrow
U^{\dag}\equiv U^{-1}$. This implies $S\rightarrow S^{\dag}$
resulting
\begin{equation}\label{inverse scattering}
Ua_i^{\dag}U^{\dag}=\sum^2_{j=1}S_{ji}a_j^{\dag}.
\end{equation}
If we substitute Eq. (\ref{inverse scattering}) into Eq.
(\ref{single photon output}) and use the scattering matrix given
by Eq. (\ref{quantum transform}), we find
\begin{equation}\label{single photon output 2}
U|10\rangle=-\sin(\phi/2)|10\rangle+\cos(\phi/2)|01\rangle,
\end{equation}
up to an overall phase. Similarly we can calculate
\begin{equation}\label{double photon output}
U|11\rangle=\frac{1}{\sqrt{2}}\sin(\phi)(-|20\rangle+|02\rangle)+\cos(\phi)|11\rangle,
\end{equation}
where the input is a pair of photons, one at each of the ports 1
and 2.

\subsection{Single-photon input vs. two-photon input}
%{\bf Single-photon input vs. two-photon input.}
The Heisenberg
picture is convenient for computing the expectation values of
photon numbers. For the single photon input $|10\rangle$, the
intensities at the detectors $D_1$ and $D_2$ reads
\begin{eqnarray}
I_1\equiv\langle b^{\dag}_1b_1\rangle=\sin^2(\phi/2),\label{single photon intensity1}\\
I_2\equiv\langle b^{\dag}_2b_2\rangle=\cos^2(\phi/2),\label{single
photon intensity2}
\end{eqnarray}
whereas for the two-photon input $|11\rangle$, we have the
single-photon counts at each detector $\langle
b^{\dag}_1b_1\rangle=1=\langle b^{\dag}_2b_2\rangle$, i.e. there
is no interference. On the other hand, by using Eq. (\ref{quantum
transform}), we can calculate the two-photon coincidences at the
detectors $D_1$ and $D_2$,
\begin{equation}\label{two-photon coincidence}
I_{12}\equiv\langle b_1^{\dag}b_2^{\dag}b_2
b_1\rangle=\cos^2(\phi),
\end{equation}
which has twofold increase in the fringe pattern. This is also
clear from the Schr\"odinger evolution of the state given by Eq.
(\ref{double photon output}). The reason of this two-fold increase
lies in the fact that when the two photons, one from each input
port, enter into the loop, they transform into the following
two-photon path-entangled state,
\begin{equation}\label{two-photon path entangled state}
|11\rangle\rightarrow\frac{|20\rangle+|02\rangle}{\sqrt{2}}\rightarrow
\frac{|20\rangle+e^{i2\phi}|02\rangle}{\sqrt{2}},
\end{equation}
which shows a two-fold reduction in the wavelength of source
photons. This was nicely demonstrated in the experiment
\cite{Edamatsu 2002} using photon pairs (biphotons) generated by
spontaneous PDC.
%\begin{figure*}
% \scalebox{0.50}{\includegraphics{Fig4}}
% \caption{(a) The plot of two-photon coincidence counts defined by the Eq. (\ref{pdc two-photon
% coincidence}), (b) the visibility of two-photon counts defined by the Eq.
% (\ref{two-photon visibility}).}
% \label{Fig4}
% \end{figure*}

\subsection{Entangled photon pairs}
%{\bf Entangled photon pairs input.}
We now discuss how the results given by Eqs. (\ref{intensity1})
and (\ref{intensity2}) are modified under conditions of arbitrary
pumping, if we work with down-converted photons. To avoid coupling
losses to a fiber, entangled photon pairs can also be generated
inside a fiber at telecom wavelengths. This has been demonstrated
by Kumar and coworkers \cite{Kumar 2005}. The state produced in
nondegenerate parametric down conversion can be written as
\begin{equation}\label{pdc state}
|\psi\rangle=\frac{1}{\cosh
r}\sum_{n=0}^{\infty}(-e^{i\theta}\tanh r)^n |n\rangle_1
|n\rangle_2.
\end{equation}
This state can be generated mathematically by applying the
two-mode squeeze operator $\hat{S}=\exp(\rho^*a_1a_2-\rho
a_1^{\dag}a_2^{\dag})$ on vacuum, where $\rho=re^{i\theta}$ is the
complex interaction parameter (also known as the squeezing
parameter) proportional to the nonlinearity of the crystal, the
pump amplitude and the crystal length. Polarization of photon
pairs in the noncollinear type-I and the collinear type-II
down-conversion are the same and orthogonal respectively. At first
glance, it is easy to see that this state is nonseparable
($\equiv$ entangled) to a product of states of mode 1 and mode 2.
It is already in Schmidt-decomposed form with a Schmidt number
higher than 1 for $r>0$, which is a measure of entanglement
\cite{Nielsen 2000}. Moreover one can calculate the {\em Entropy
of Entanglement} \cite{Bennett 1996}, $E=-{\mathrm
Tr}_2\rho\log_2\rho$ as a function of $r$,
\begin{equation}\label{entropy of entanglement}
E=\cosh^2r\log_2(\cosh^2r)-\sinh^2r\log_2(\sinh^2r).
\end{equation}
The amount of entanglement given by Eq. (\ref{entropy of
entanglement}) is approximately linear in $r$ showing that the
state in Eq. (\ref{pdc state}) is fully entangled for
$r\rightarrow\infty$.

When the input modes are in the state given by the Eq. (\ref{pdc
state}) the output detectors $D_1$ and $D_2$ read the single
counts
\begin{equation}\label{pdc intensity output}
I_1\equiv\langle b^{\dag}_1b_1\rangle=\sinh^2r=\langle
b^{\dag}_2b_2\rangle\equiv I_2,
\end{equation}
which does not give any information on the rotation. On the other
hand, the two-photon coincidence counts, after subtracting
independent counts, normalized over the product of single counts
at each detector becomes,
\begin{equation}\label{pdc two-photon coincidence}
g^2_{12}=\frac{\langle b_1^{\dag}b_2^{\dag}b_2 b_1\rangle}{\langle
b^{\dag}_1b_1\rangle\langle
b^{\dag}_2b_2\rangle}-1=\cos^2(\phi)\coth^2r,
\end{equation}
% \begin{figure}
% \scalebox{0.53}{\includegraphics{Fig5}}% Here is how to import EPS art
% \caption{\label{Fig5}The visibility of two-photon counts defined by the Eq.
% (\ref{two-photon visibility}).}
% \end{figure}
% \begin{figure}[h]
% \scalebox{0.70}{\includegraphics{Fig5}}% Here is how to import EPS art
% \caption{\label{Fig5}Four-photon coincidence counts defined in Eq. (\ref{4photon in detector})
% with different interaction parameter values.}
% \end{figure}
which is twice more sensitive to the rotation than the result in
Eq. (\ref{single photon intensity1}) with 100\% visibility. The
signal itself depends on $r$ and it is most significant when
$r\approx 1$. The regime with an interaction parameter value of
$r=1.39$ has already been reached in the experiment \cite{Caminati
2006}.
%The plot of this quantity with different values of $r$ is shown in Fig. \ref{Fig4}(a). The
%visibility of the two-photon counts,
%\begin{equation}\label{two-photon visibility}
%V=\frac{1}{1+2\tanh^2r},
%\end{equation}
%with maximum 1 for small $r$ and minimum $1/3$ for
%$r\rightarrow\infty$, is shown in Fig. \ref{Fig4}(b).
%
%The reason of the constant term appearing in the Eq. (\ref{pdc
%two-photon coincidence}) is the coincidences coming from
%independent photon counts at the detectors. Since these accidental
%coincidences ($I_1\cdot I_2=\sinh^4r$) are insensitive to the
%phase $``\phi"$ it reduces the visibility acting as a background
%noise especially for large values of $r$. Thus, one way of
%increasing the visibility would be to subtract the accidental
%coincidences from the signal.

On the other hand, we can employ a different measurement --the
projective measurement,
\begin{eqnarray}\label{two-photon probability}
P_2&=&{\mathrm
Tr}[|11\rangle\langle11|\rho(\phi)]\equiv|\langle11|U|\psi\rangle|^2\nonumber\\
&=&\frac{\tanh^2r}{\cosh^2r}\cos^2(\phi),
\end{eqnarray}
which is the probability of detecting two photons, one at each
detector operating in coincidence. Here the state
$\rho(\phi)=U|\psi\rangle\langle\psi|U^{\dag}$ is the evolved
density matrix corresponding to the state given in  Eq. (\ref{pdc
state}). This probability can be easily calculated by utilizing
the Schr\"odinger picture evolution of the state vector
$|11\rangle$ given in  Eq. (\ref{double photon output}). The
expression in Eq. (\ref{two-photon probability}) is a pure
two-photon interference effect showing by halving the de Broglie
wavelength of the source photons. Thus, the two-photon
coincidences by using the state in Eq. (\ref{pdc state}) shows
two-fold increase in the sensitivity of the phase measurement.

%On the other hand one can also calculate the coincidence counts of
%four photons two-by-two at each detector as given by Glauber's
%higher order correlation functions:
%\begin{widetext}
%\begin{eqnarray}\label{4photon in
%detector} \langle {b_1}^{\dag 2}{b_2}^{\dag
%2}{b_2}^2{b_1}^2\rangle =(3\cos^2(\phi)-1)^2\sinh^4r\cosh^4r
%+4(3\cos^2(\phi)+1)\sinh^6r\cosh^2r+4\sinh^8r.
%\end{eqnarray}
%\end{widetext}
 \begin{figure}
 \centering
 \scalebox{.8}{\includegraphics{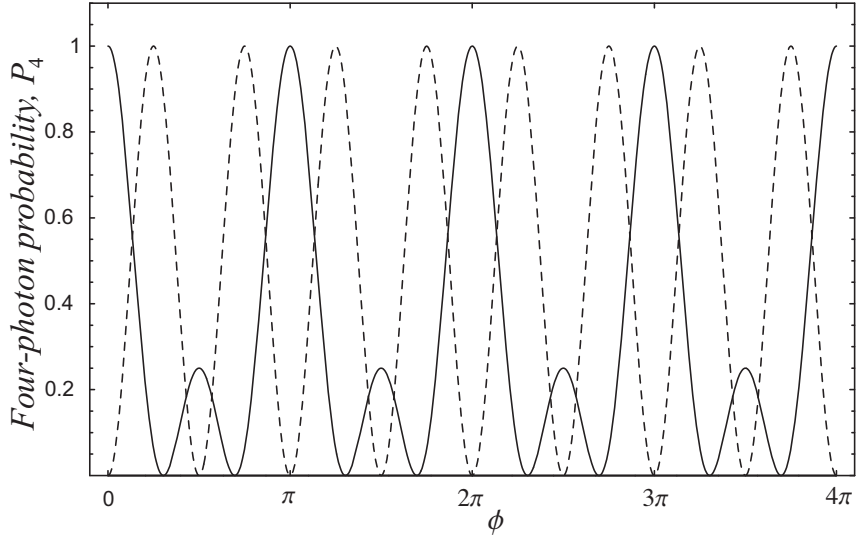}}% Here is how to import EPS art
 \caption{\label{Fig4}Normalized four-photon probability in coincidence in 2-by-2 (solid line) and  3-by-1 (dashed line)
 detection scheme described by  Eqs. (\ref{four-photon probability}) and (\ref{four-photon probability 3by1}) respectively.}
 \end{figure}
The next question is-- can we increase the sensitivity further by
measuring higher order coincidences? We suggest employment of four
single-photon detectors $D_i$ ($i=1,2,3,4$) as depicted in Fig.
\ref{Fig3}. We note that in a recent experiment \cite{Ourjoumtsev
2006} the tomography of the Fock state $|2\rangle$ was done by
letting the two photons propagate in different directions and by
single photon detectors. For detection of multi-photons, it is
easier to use single photon detectors. We examine the coincidence
of clicking four detectors i.e. the probability of the state
$|1_{D_1}1_{D_3}1_{D_2}1_{D_4}\rangle$ where $D_i$'s denote the
modes that goes into the corresponding detectors. This requires
the state in modes $b_1$ and $b_2$ before the beam splitters
$BS_1$ and $BS_2$ to be in a four-photon subspace. We now outline
this calculation. The four-photon coincidence detection
probability is given by
\begin{eqnarray}\label{four-photon probability}
P_4&=&|\langle1_{D_1}1_{D_3}1_{D_2}1_{D_4}|U_BU_S|\psi,0_{v_1}0_{v_2}\rangle|^2\nonumber\\
&=&\langle1_{D_1}1_{D_3}1_{D_2}1_{D_4}|U_BU_S|\psi,0_{v_1}0_{v_2}\rangle\langle
\psi,0_{v_1}0_{v_2}|U_S^{\dag}U_B^{\dag}|1_{D_1}1_{D_3}1_{D_2}1_{D_4}\rangle,
\end{eqnarray}
where the state $|\psi,0_{v_1}0_{v_2}\rangle$ represents the
tensor product of the state (\ref{pdc state}) with the vacuum
ports $v_1$ and $v_2$ at the beam splitters $BS_1$ and $BS_2$. The
unitary operators $U_S$ and $U_B$ represent the evolution of the
states inside the Sagnac interferometer and the two beam splitters
$BS_1$ and $BS_2$ respectively. First, we begin by calculating the
inverse evolution
\begin{eqnarray}\label{calc1}
U_B^{\dag}|1_{D_1}1_{D_3}1_{D_2}1_{D_4}\rangle&=&U_B^{\dag}D_1^{\dag}D_3^{\dag}D_2^{\dag}D_4^{\dag}U_BU_B^{\dag}|0000\rangle\nonumber\\
&=&(t_1^*b_1^{\dag}+r_1'^*v_1^{\dag})(t_2^*b_2^{\dag}+r_2'^*v_2^{\dag})(r_1^*b_1^{\dag}+t_1'^*v_1^{\dag})(r_2^*b_2^{\dag}+t_2'^*v_2^{\dag})|0000\rangle\nonumber\\
&=&t_1^*t_2^*r_1^*r_2^*b_1^{\dag2}b_2^{\dag2}|0000\rangle+\ldots,
\end{eqnarray}
where we take only four-photon state in modes $b_1$ and $b_2$
because other terms are irrelevant in our calculation. Here
$t_i$'s and $r_i$'s are transmittance and reflectance coefficients
of the beam splitter $BS_i$ ($i=1,2$). Next, we operate
$U_S^{\dag}$ on the resultant state above,
\begin{eqnarray}\label{calc2}
U_S^{\dag}t_1^*t_2^*r_1^*r_2^*b_1^{\dag2}b_2^{\dag2}|0000\rangle&=&t_1^*t_2^*r_1^*r_2^*U_S^{\dag}b_1^{\dag2}b_2^{\dag2}U_SU_S^{\dag}|0000\rangle\nonumber\\
&=&t_1^*t_2^*r_1^*r_2^*\left(-\sin(\phi/2)a_1^{\dag}+\cos(\phi/2)a_2^{\dag}\right)^2\nonumber\\
&\times&\!\!\!\!\!\!\!\!\left(\cos(\phi/2)a_1^{\dag}+\sin(\phi/2)a_2^{\dag}\right)^2|0000\rangle\nonumber\\
&=&\left(\frac{1}{2}\sin(\phi)(-a_1^{\dag2}+a_2^{\dag2})+\cos(\phi)a_1^{\dag}a_2^{\dag}\right)^2|0000\rangle\nonumber\\
&=&\frac{1}{2}[1+3\cos(2\phi)]|2200\rangle+\ldots,
\end{eqnarray}
where we use the transformation given by Eq. (\ref{quantum
transform}) between the modes $a_1$, $a_2$ and $b_1$, $b_2$. In
the last line of the Eq. (\ref{calc2}) the first two modes are
$a_1$ and $a_2$, while the last two modes are the vacuum modes of
the beam splitters $BS_1$ and $BS_2$. In the last line of Eq.
(\ref{calc2}) we take only the state which has equal number of
photons in modes $a_1$ and $a_2$ because the input state
$|\psi\rangle$ is a pair photon state which is given in Eq.
(\ref{pdc state}). Therefore the absolute square of the inner
product of the resultant state given in the Eq. (\ref{calc2}) with
$|\psi,0_{v_1}0_{v_2}\rangle$ gives us the four-photon coincidence
probability
\begin{equation}\label{four-photon probability}
P_4=\frac{\tanh^4r}{\cosh^2r}|t_1t_2r_1r_2|^2\frac{1}{4}[1+3\cos(2\phi)]^2.
\end{equation}
\begin{figure}
  \centering
  \scalebox{1.7}{\includegraphics{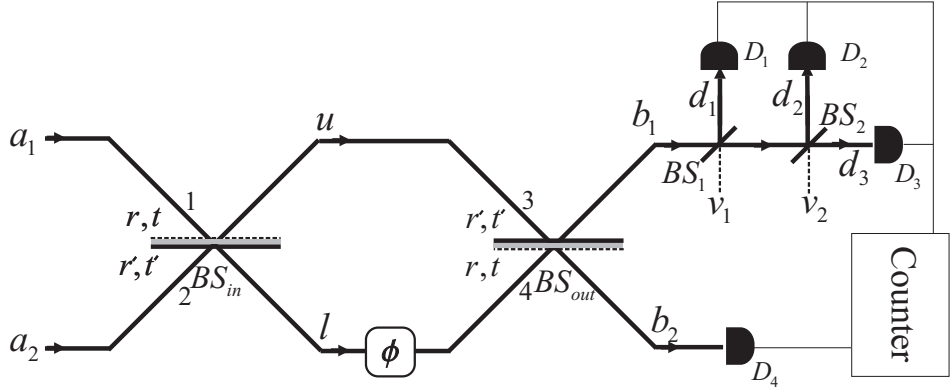}}
  \caption{The Sagnac interferometer setup for four-photon coincidence detection in 3-by-1 scheme.}
  \label{Fig5}
\end{figure}
The result given by Eq. (\ref{four-photon probability}) shows a
reduction in the period of fringes by developing smaller peaks as
depicted in Fig. \ref{Fig4}. The phase sensitivity shows a
four-fold increase w.r.t. the result in Eq. (\ref{single photon
intensity1}) obtained by single-photon input.

The four-photon coincidence detection can be done in an
alternative way as depicted in figure \ref{Fig5}. Here, three of
the four photons are to be detected in the upper output channel
$b_1$ while the fourth one goes into the detector placed in the
lower output channel $b_2$. This time, we place the two extra beam
splitters $BS_1$ and $BS_2$ to split up the three photons into
single photons before they arrive at the detectors $D_1$, $D_2$
and $D_3$. Now, we begin with the evolution of the four-photon
subspace term $|22\rangle$ provided by the input state
$|\psi\rangle$,
\begin{eqnarray}\label{3-1calc}
|22\rangle &=&
\frac{a_1^{\dag2}}{\sqrt{2}}\frac{a_2^{\dag2}}{\sqrt{2}}|00\rangle\,\,\longrightarrow\,\,\frac{1}{2}
\left(\frac{iu^{\dag}+l^{\dag}}{\sqrt{2}}\right)^2\left(\frac{u^{\dag}+il^{\dag}}{\sqrt{2}}\right)^2|00\rangle\nonumber\\
%&=&\frac{1}{8}(u^{\dag4}+l^{\dag4}-2u^{\dag2}l^{\dag2})|00\rangle\nonumber\\
&\longrightarrow&-\frac{1}{8}
\left\{\left(\frac{ib_1^{\dag}+b_2^{\dag}}{\sqrt{2}}\right)^4+e^{i4\phi}\left(\frac{b_1^{\dag}+ib_2^{\dag}}{\sqrt{2}}
\right)^4
+2e^{i2\phi}\left(\frac{ib_1^{\dag}+b_2^{\dag}}{\sqrt{2}}\right)^2\left(\frac{b_1^{\dag}+ib_2^{\dag}}{\sqrt{2}}\right)^2\right\}|00\rangle\nonumber\\
&=&\frac{1}{4}\left\{e^{i2\phi}\sin(2\phi)b_1^{\dag3}b_2^{\dag}+\cdots\right\}|00\rangle\nonumber\\
&\longrightarrow&\frac{e^{i2\phi}}{4}\sin(2\phi)\left\{\left(r_1d_1^{\dag}+t_1(r_2d_2^{\dag}+t_2d_3^{\dag})\right)^3b_2^{\dag}\right\}|0000\rangle\nonumber\\
&=&\frac{e^{i2\phi}}{4}\sin(2\phi)\left\{6r_1t_1^2r_2t_2d_1^{\dag}d_2^{\dag}d_3^{\dag}b_2^{\dag}+\cdots\right\}|0000\rangle\nonumber\\
&=&\frac{3}{2}e^{i2\phi}\sin(2\phi)r_1t_1^2r_2t_2|1111\rangle.
\end{eqnarray}
The  arrow in the first line represents the evolution of the input
modes into the interferometer after $BS_{in}$, while the second
arrow shows  further evolution of the modes by the phase shift
$\phi$  and  $BS_{out}$. In the third line we omit the terms
giving photons in the channels other than three in $b_1$ and one
in $b_2$. The fourth line shows how the channel $b_1$ split up
into the modes $d_1$, $d_2$ and $d_3$ going into the detectors
$D_1$, $D_2$ and $D_3$ respectively and since we are considering
only one photon per detector we omit the other terms in the
following line. In the last line we obtain the state corresponding
to the four-photon coincidence detection in 3-by-1 scheme. Then,
we calculate the probability of four-photon coincidence to be,
\begin{equation}\label{four-photon probability 3by1}
P_4^{(3by1)}=\frac{\tanh^4r}{\cosh^2r}|t_1^2t_2r_1r_2|^2\frac{9}{4}\sin^2(2\phi).
\end{equation}
The normalized plot of this probability is shown in Fig.
\ref{Fig4}. Note that all the peaks are even in the interference
pattern showing a pure four-fold increase in the sensitivity.
 We note that Steuernagel  has a similar result in his work
\cite{Steuernagel 2002} on reduced de Broglie wavelength using
precisely two photon events at each detector. In our proposal
above we use single photon detectors to do four-photon
coincidence. Currently efforts are on to find efficient nonlinear
absorbers so that these could be used for detection of the precise
number of photons \cite{Korobkin 2002,Agarwal 2006, Chang 2006}.

 %If all the peaks had
%been even, then this would have been a pure four-photon
%interference effect, because the four-photon coincidence projects
%the state (\ref{pdc state}) into the four-photon subspace
%$|22\rangle$, and this becomes
%\begin{equation}\label{four-photon inside the interferometer}
%|22\rangle\rightarrow\sqrt{\frac{3}{4}}\left(\frac{|40\rangle+e^{i4\phi}|04\rangle}{\sqrt{2}}\right)
%+\frac{1}{\sqrt{4}}e^{i2\phi}|22\rangle,
%\end{equation}
%inside the interferometer. The first state in the parenthesis on
%the r.h.s leads to the pure four-photon interference in $75\%$ of
%all cases. Only one quarter of all cases the unwanted state
%$|22\rangle$ occurs.

\section{Conclusion}\label{conclusion}

%We showed that the use of parametric down conversion in Sagnac
%interferometer increases the sensitivity to rotations by a factor
%of two or four in comparison to classical or single-photon light
%depending on the detection mechanism.
There are advantages of using single photon interferometer as then
the unwanted effects due to nonlinearities are avoided. However
the integration time becomes long so that one can achieve the same
level of sensitivity as classical interferometers \cite{Lefevre
1993}. What we are demonstrating is that we get superresolution
relative to what is obtained by the usage of single photons. We
think that experiments should be feasible because many two-photon
and four-photon interference effects have been observed \cite{Ou
1990a, Rarity 1990, Ou 1990b, Edamatsu 2002, Eibl 2003, Walther
2004}.

There is one thing that should be in consideration in the use
fiber. All fiber ring interferometers make use of single-mode
fibers. Normally all single-mode fibers permit the transmission of
modes of two orthogonal polarizations through the fiber in both
directions. Disturbances, such as temperature fluctuations and
mechanical stresses introduces birefringence to the fiber causing
one mode to be transferred to the other. The noise produced by the
transfer of modes from one to the other may effect the
interference pattern. However by utilizing half-wave plates and
polarization controllers \cite{Kumar 2005} this noise may be
suppressed.

We are grateful to NSF grant no CCF 0524673 for supporting this
research.

%Reply to the Editor
%
%We first all thank to referees for their useful suggestions. We
%are submitting the revised version. We carried out all the changes
%according to the referee 1 and the ones of the referee 2 except
%one for reasons mentioned below. We have also taken the
%opportunity to expand the paper a bit. We have added appropriate
%comments at various places.
%
%We do not discuss the propagation characteristics of the entangled
%photons in a fiber as at the intensities we work the fiber is a
%linear medium and all is very well understood in a linear medium.
%Further the work of Kumar and coworkers concern the propagation of
%a strong field in a Kerr medium  and the generation of entangled
%photons. This is not what we do. Their group addresses a different
%problem. Though we are happy to add in the manuscript a comment
%that their scheme for production of entangled photons can also be
%utilized. Brief comments about resolution is added in conclusions.


\begin{thebibliography}{99}

 \bibitem{Sagnac 1913a}
   G. Sagnac, ``L'ether lumineux demontre par l'effect du vent relatif d'ether dans un
  interferometre en rotation uniforme," C. R. Acad. Sci. {\bf 157,} 708--710 (1913).

% \bibitem{Post 1967}
% E. J. Post, Rev. Mod. Phys. {\bf 39}, 475
% (1967).

  \bibitem{Bertocchi 2006}
  G. Bertocchi,  O. Alibart,  D. B. Ostrowsky,  S. Tanzilli, and
  P. Baldi, ``Single-photon Sagnac interferometer," J. Phys. B {\bf 39,}
 1011--1016 (2006).

 \bibitem{Grangier 1986}
  P. Grangier,  G. Roger, and  A. Aspect, ``Experimental evidence for a photon anticorrelation effect on a
 beam splitter: a new light on single-photon interferences," Europhys. Lett. {\bf 1,}  173
 (1986).


 \bibitem{Hariharan 1993}
  P. Hariharan,  N. Brown, and  B. C. Sanders, ``Interference of independent laser-beams at
 the single-photon level," \jmo {\bf 40,} 113--122 (1993).


 \bibitem{Zeilinger 2000}
  A. Zeilinger, ``Experiment and the foundations of quantum
 physics," Rev. Mod. Phys. {\bf 71,} S288--S297 (1999).

% \bibitem{Caves 1981}
% Caves C M, ``Quantum-mechanical noise in an interferometer," \prd {\bf 23,} 1693--1708
% (1981).
%
% \bibitem{Yurke 1986}
% Yurke B, McCall S L, and Klauder J R, ``SU(2) and SU(1,1) interferometers," \pra  {\bf 33,}
% 4033--4054 (1986).

 \bibitem{Holland 1993}
  M. J. Holland and  K. Burnett, ``Interferometric detection of optical-phase shifts at the
  heisenberg limit,"  \prl {\bf 71,} 1355--1358 (1993).

 \bibitem{Dowling 1998}
  J. P. Dowling, ``Correlated input-port, matter-wave interferometer: Quantum-noise limits
 to the atom-laser gyroscope,"  \pra {\bf 57,} 4736-4746 (1998).

 \bibitem{Abouraddy 2001}
  A. F. Abouraddy,  B. E. A. Saleh,  A. V. Sergienko, and  M. C. Teich, `` Role of entanglement in two-photon imaging ,"
 \prl {\bf 87,} 123602 (2001).

 \bibitem{Pittman 1995}
  T. B. Pittman,  Y. H. Shih,  D. V. Strekalov, and  A. V. Sergienko, ``Optical imaging by means of 2-photon
 quantum entanglement," \pra {\bf 52,} R3429--R3432 (1995).

 \bibitem{Boto 2000}
  A. N. Boto,  P. Kok,  D. S. Abrams,  S. L. Braunstein,  C. P. Williams,
 and  J. P. Dowling, ``Quantum interferometric optical lithography: Exploiting entanglement
 to beat the diffraction limit,"  \prl {\bf 85,} 2733--2736 (2000).

 \bibitem{Yablonovitch 1999}
 E. Yablonovitch and R. B. Vrijen, ``Optical projection
 lithography at half the Rayleigh resolution limit by two-photon
 exposure," Opt. Eng. {\bf 38}, 334--338 (1999).

 \bibitem{Korobkin 2002}
 D. V. Korobkin and E. Yablonovitch, ``Two-fold spatial resolution
 enhancement by two-photon exposure of photographic film," Opt.
 Eng. {\bf 41}, 1729--1732 (2002).

 \bibitem{Agarwal 2001}
  G. S. Agarwal,  R. W. Boyd,  E. M. Nagasako, and  S. J. Bentley, ``Comment on "Quantum interferometric optical lithography:
 Exploiting entanglement to beat the diffraction limit","  \prl {\bf
 86,} 1389--1389 (2001).

 \bibitem{Bjork 2001}
  G. Bj\"ork,  L. L. Sanchez-Soto, and  J. S\"oderholm, ``Entangled-state lithography:
 Tailoring any pattern with a single state,"  \prl {\bf 86,}
 4516--4519 (2001).

 \bibitem{Dangelo 2001}
  M. D'Angelo,  M. V. Chekhova, and  Y. Shih, ``Two-photon diffraction and quantum lithography,"  \prl {\bf 87,} 013602
 (2001).

 \bibitem{Agarwal 2003}
  G. S. Agarwal and  M. O. Scully, ``Magneto-optical spectroscopy with entangled photons,"  \ol {\bf 28,}
 462--464 (2003).

 \bibitem{Ou 1990a}
  Z. Y. Ou,  L. J. Wang,  X. Y. Zou, and   L. Mandel,``Evidence for phase memory in 2-photon down conversion
 through entanglement with the vacuum ," \pra {\bf 41,}
 566--568 (1990).

 \bibitem{Rarity 1990}
  J. G. Rarity,  P. R. Tapster,  E. Jakeman,  T. Larchuk,  R. A. Campos,  M. C. Teich, and  B. E. A. Saleh,
 ``2-photon interference in a mach-zehnder interferometer,"  \prl {\bf 65,}
 1348--1351 (1990).

 \bibitem{Ou 1990b}
  Z. Y. Ou,  X. Y. Zou,  L. J. Wang, and  L. Mandel, ``Experiment on nonclassical 4th-order interference ,"  \pra {\bf 42,}
 2957--2965 (1990).


 \bibitem{Edamatsu 2002}
  K. Edamatsu,  R. Shimizu, and  T. Itoh, ``Measurement of the photonic de Broglie wavelength of entangled
 photon pairs generated by spontaneous parametric down-conversion ,"  \prl {\bf 89,} 213601
 (2002).


 \bibitem{Eibl 2003}
  M. Eibl,  S. Gaertner,  M. Bourennane,  C. Kurtsiefer,  M. Zukowski,
 and  H. Weinfurter, ``Experimental observation of four-photon entanglement from parametric down-conversion
,"  \prl {\bf 90,} 200403 (2003).

 \bibitem{Walther 2004}
  P. Walther,   J. W. Pan,   M. Aspelmeyer,   R. Ursin,  S. Gasparoni, and
  A. Zeilinger, ``De Broglie wavelength of a non-local four-photon state
," \nat {\bf 429,} 158--161 (2004).

 \bibitem{Jacobson 1995}
  J. Jacobson,  G. Bj\"ork,   I. Chuang, and  Y. Yamomoto, ``Photonic de broglie waves,"  \prl {\bf
 74,} 4835--4838 (1995).

 \bibitem{Steuernagel 2002}
O. Steuernagel, ``de Broglie wavelength reduction for multiphoton
wave packet," \pra {\bf 65,} 033820 (2002).

 \bibitem{Schleich 1984}
  W. Schleich and  M. O. Scully 1984 {\it Modern Trends in Atomic and Molecular Physics, Proceedings of
 Les Houches Summer School, Session XXXVIII}, edited by R. Stora
 and G. Grynberg, North Holland, Amsterdam

 \bibitem{Post 1967}
  E. J. Post, ``Sagnac effect,"  \rmp {\bf 39,} 475 (1967).

 \bibitem{Jacobs 1982}
   F. Jacobs and   R. Zamoni, ``Laser ring gyro of arbitrary shape and rotation axis ,"
  Am. J. Phys. {\bf 50,} 659--660 (1982).

 \bibitem{Kumar 2005}
  X. Li,  P. L. Voss,  J. E. Sharping, and  P. Kumar, ``Optical-fiber source of polarization-entangled photons in the 1550 nm telecom
 band," \prl {\bf 94,} 053601 (2005).

 \bibitem{Nielsen 2000}
  M. A. Nielsen and   I. L. Chuang,  {\em Quantum computation and quantum
 information}, (Cambridge University Press, Cambridge, 2000).

 \bibitem{Bennett 1996}
  C. H. Bennett,  D. P. DiVincenzo,  J. A. Smolin, and   W. K. Wooters, ``Mixed-state entanglement and quantum
 error correction,"
 {\it Phys. Rev.} A {\bf 54,} 3824--3851 (1996).

 \bibitem{Caminati 2006}
  M. Caminati,   F. De Martini,   R. Perris,   F. Sciarrino, and
  V. Secondi, ``Nonseparable Werner states in spontaneous parametric down-conversion,"  {\it Phys. Rev.} A {\bf 73,} 032312 (2006).

 \bibitem{Ourjoumtsev 2006}
  A. Ourjoumtsev,  R. Tualle-Brouri,  and P. Grangier,``Quantum homodyne tomography of a two-photon fock state
 ," \prl {\bf 96,} 213601 (2006).

 \bibitem{Agarwal 2006}
 G. S. Agarwal, K. W. Chan, R. W. Boyd, H. Cable, and J. P. Dowling
 ``Quantum states of light produced by
 a high-gain optical parametric amplifier for use in quantum
 lithography," \josab {\bf 24,} 270 (2007).

 \bibitem{Chang 2006}
 H. J. Chang, H. Shin, M. N. O'Sullivan-Hale, and R. W. Boyd, ``Implementation of subRayleigh
 lithography using an $N$-photon absorber," J. Mod. Opt. {\bf 53,}
 2271-2277 (2006).

 \bibitem{Lefevre 1993}
  H. Lefevre, {\em The Fiber-Optic Gyroscope}, (Artech House, Boston, 1993).

\end{thebibliography}
 \end{document}